\documentclass[12pt]{article}
\usepackage{epsfig}
\usepackage{amssymb}
\usepackage{amsmath}
\newcommand{\singlespace}{\renewcommand{\baselinestretch}{1}\large\normalsize}
\newcommand{\doublespace}{\renewcommand{\baselinestretch}{1.6}\large\normalsize}
\newcommand{\beq}{\begin{equation}}
\newcommand{\eeq}{\end{equation}}
\newcommand{\bea}{\begin{eqnarray}}
\newcommand{\eea}{\end{eqnarray}}

\newcommand{\dslash}{\partial\!\!\!/}

\newcommand{\pb}{\bar\psi}

\def\roughly#1{\mathrel{\raise.3ex\hbox{$#1$\kern-.75em%
\lower1ex\hbox{$\sim$}}}}

\def\={\;=\;}
\def\+{\;+\;}

\begin{document}

\begin{flushright}
May 2004
\end{flushright}
\vspace{0.25cm}
\begin{center}
\doublespace
\begin{large}
{\bf Quark mass effects on the stability of hybrid stars}
\end{large}
\vskip 0.5in
M. Buballa$^{a}$,
F. Neumann$^{b}$,
M. Oertel$^{c}$, 
and
I. Shovkovy$^d$ \\[0.5cm]
{\small{\it $^a$Institut f\"ur Kernphysik, TU Darmstadt,
                Schlossgartenstr. 9, 64289 Darmstadt, Germany\\
            $^b$Inst. f. Materialwissenschaften, TU Darmstadt,
                Petersenstr. 23, 64287 Darmstadt, Germany\\
            $^c$CEA Bruy\`eres le Ch\^atel, DPTA/SPN, BP 12, 91680 Bruy\`eres
                le Ch\^atel, France\\
            $^d$Institut f\"ur Theoretische Physik, Universit\"at Frankfurt,
                60054 Frankfurt/Main, Germany}} 
\end{center}
\vspace{0.75cm}

\begin{abstract}
We perform a study of the possible existence of hybrid stars with color
superconducting quark cores using a specific hadronic
model in a combination with an NJL-type quark model.
It is shown that the constituent mass of the non-strange
quarks in vacuum is a very important parameter
that controls the beginning of the hadron-quark phase transition.
At relatively small values of the mass, the first quark phase
that appears is the two-flavor
color superconducting (2SC) phase which, at larger densities, 
is replaced by the color-flavor locked (CFL) phase. At large 
values of the mass, on the other hand, the phase transition 
goes from the hadronic phase directly into the CFL phase 
avoiding the 2SC phase. It appears, however, that the only stable 
hybrid stars obtained are those with the 2SC 
quark cores. 
\end{abstract}
\vspace{0.5cm}
\singlespace

\newpage

\section{Introduction}

At low temperatures and high densities, strongly interacting matter is
expected to be a color superconductor.
The prediction that the related energy gaps could be of the order of
100~MeV in quark matter at densities of a few times
nuclear matter density~\cite{ARW98,RSSV98} evoked many theoretical studies 
of color superconducting matter at intermediate and high densities,
suggesting that the structure of the QCD phase diagram could be very rich
(for reviews see, e.g., Refs.~\cite{RaWi00, Alford01,Schaefer03,Rischke03}).
Most prominent are the two-flavor color superconducting (2SC) phase 
where only up and down quarks are paired and the color-flavor locked
(CFL) phase where up, down and strange quarks participate in a
diquark condensate. 

Since the temperatures attained in heavy-ion collisions are probably 
too high to produce color superconducting phases, compact stars 
seem to be the most promising objects to test these ideas
empirically. Therefore, many theoretical investigations have been
performed to identify possible signals of color superconducting
quark matter in the interior of stars. At this point, obviously 
the first question is whether the equation of state of strongly 
interacting matter allows for the existence of quark cores inside
stars (which should be called hybrid stars then). This 
question has already been addressed in several works using 
different approaches with taking the effects of 
color superconductivity into account (see, for example, 
Refs.~\cite{AlRe03,BB02,BBBNOS02,BGAYT,SHH03}). 

Since it is a very difficult task to construct an equation of state
for hadronic and quark matter on the same footing, usually a hybrid
equation of state is employed. This consists in taking some realistic
hadronic equation of state and combining it with some description of
deconfined quark matter. The phase transition point is thereby
determined using the criterion of maximal pressure.  Studies employing
a bag-model description of quark-matter phases often show reasonable
windows in parameter space where hybrid stars could exist. Moreover,
bag-model studies revealed that a diquark condensation in color
superconducting phases, although it contributes only to subleading order 
to the pressure, could have a large
effect on the hadron-quark phase transition~\cite{LH02} and on the
properties of compact stars~\cite{AlRe03} because the bag constant
cancels out most of the leading $\mu^4$ term. On the other hand,
bag-model calculations miss possible effects of the density and phase
dependence of the quark masses, diquark gaps and the bag constant.
Several authors have therefore employed NJL-type quark models where
these effects can be studied.

Two early investigations in this direction were performed in
Refs.~\cite{SLS99,SPL00}. It was shown that quarks can possibly
exist in a hadron-quark mixed phase inside a star, but not in a pure
quark core. This result could be traced back to the 
effective bag constant and the effective strange quark mass, 
which are dynamically generated quantities in NJL-type models and 
which are usually quite large.
These investigations, however, did not include the possibility
of diquark condensation. By including such effects, the authors 
of Refs.~\cite{BGAYT,SHH03} found stable hybrid stars with pure 
quark cores using an NJL description of the quark phase. This is
in contrast to the findings of Ref.~\cite{BBBNOS02} where no stable
configurations containing a pure quark core were found even if
color superconductivity was taken into account. 

As our knowledge based on first principles about the quark phase at
relevant densities is rather limited and we have to rely on model
studies, it is very important to sort out
the physical reasons for these, at first sight contradictory,
findings. We will thereby concentrate on the differences between the
approaches of Ref.~\cite{BBBNOS02} and Ref.~\cite{SHH03}, since this
already suffices to show the main point.  The most obvious difference
between these two studies is the fact that the quark-matter equation
of state has been derived within a two-flavor model in
Ref.~\cite{SHH03} and within a three-flavor model in
Ref.~\cite{BBBNOS02}. In fact, it was found by the latter that the
hadron-quark phase transition takes place only if strange quarks play
a non-negligible role in quark matter. In particular, if diquark
condensates were included, the transition always went directly to the
CFL phase, whereas the 2SC phase did not appear.  This is in
qualitative agreement with earlier arguments given in Ref.~\cite{AR02}
where it was claimed that the 2SC phase is never favored if electric
and color neutrality is imposed, as required for compact stars.
However, since the strange quarks are relatively heavy, it was found
in Ref.~\cite{BBBNOS02} that their emergence causes a strong increase
of the energy density at the phase transition point, rendering the
would-be hybrid star unstable against collapse.

This problem did not arise in Ref.~\cite{SHH03} where strange quarks 
have not been taken into account. Here the energy density increases 
only slightly due to the quark degrees of freedom and the resulting 
hybrid equation of state can accommodate stable compact stars with 
pure quark-matter cores. However, in view of Ref.~\cite{BBBNOS02}, 
one may ask whether this result would persist if strange quarks were 
added to this model. One may also wonder, how a hadron-quark phase 
transition in Ref.~\cite{SHH03} without strange quarks was possible 
at all.

Let us begin with the second question. 
A priori, there are two possible sources of the observed differences: 
the hadronic part and the quark part of the equation of state.
Since our firm knowledge about the hadronic part is
limited to the vicinity of the nuclear matter saturation point,
there can be large differences in the behavior at higher densities. 
To get an idea about the variations, the authors of 
Ref.~\cite{BBBNOS02} have employed three rather different hadronic equations 
of state. Thus, although yet another hadronic equation of state has been 
used in Ref.~\cite{SHH03}, the essential difference is more likely
to come from the quark part.

In fact, if we compare the vacuum properties of the two NJL models,
we see that the constituent masses of the up and down quarks are 
$M_u^{\mathit vac} =$~368~MeV
in Ref.~\cite{BBBNOS02}, but only 314~MeV in Ref.~\cite{SHH03}.
If these were non-interacting constituent quarks, this would mean that
the quark gas reaches zero pressure only at 
$\mu_B = 3\,M_u^{\mathit vac} \simeq$~1100~MeV in the former, but already
at the nucleon mass in the latter case.  
From this point of view, it appears plausible 
that it is much harder for the heavier quarks of Ref.~\cite{BBBNOS02} to
compete with hadronic matter than for the lighter ones of Ref.~\cite{SHH03}.
In an NJL model, the quarks are of course interacting and their masses
at large densities are much smaller than in vacuum.
Nevertheless, as we will see below, the net effect is quite similar.

To investigate this point more carefully (as well as the question about
the influence of the strange quarks), we consider the following
NJL type quark model defined by the Lagrangian density
\beq 
\mathcal{ L} \= \bar{\psi} (i \dslash
- \hat{m}) \psi \+ \mathcal{L}_{q\bar q} \+ \mathcal{L}_{qq},
\label{Lagrange}
\end{equation}
where $\psi$ denotes a quark field with three flavors and three colors. 
The mass matrix $\hat m$ has the form 
$\hat m = {\rm diag}(m_u, m_d, m_s)$ in flavor space.
The interaction splits into a quark-quark part
\beq
    \mathcal{L}_{qq} \=
    H\sum_{A = 2,5,7} \sum_{A' = 2,5,7}
    (\pb \,i\gamma_5 \tau_A \lambda_{A'} \,C\pb^T)
    (\psi^T C \,i\gamma_5 \tau_A \lambda_{A'} \, \psi) 
    \,
\label{Lqq}
\eeq
and a quark-antiquark part
\bea
    \mathcal{L}_{q\bar q} &\=& G\, \sum_{a=0}^8 \Big[(\pb \tau_a\psi)^2
    \+ (\pb i\gamma_5 \tau_a \psi)^2\Big]\nonumber \\ && 
    \;-\; K\,\Big[{\rm det}_f\Big(\pb(1+\gamma_5)\psi\Big) \,+\
                   {\rm det}_f\Big(\pb(1-\gamma_5)\psi\Big)\Big]\;.
\label{Lqbarq}
\eea
Here $\tau_i$ and $\lambda_j$ are $SU(3)$- (Gell-Mann-)matrices in
flavor and color space, respectively.  $\tau_0 =
\sqrt{\frac{2}{3}}\,1\hspace{-1.5mm}1_f$ is proportional to the unit
matrix in flavor space.
The same three-flavor model has been employed in Ref.~\cite{BBBNOS02}
\footnote{For a more detailed analysis of electrically and color neutral
quark matter within this model, see also Refs.~\cite{SRP02,NBO03}}.  

To study the quark mass dependence, we use two different parameter
sets throughout the following analysis. The first one
(hereafter ``RKH'') is taken from Ref.~\cite{rehberg} and is the
parameter set employed in Ref.~\cite{BBBNOS02}, while the second set
(hereafter ``HK'') is taken from Ref.~\cite{hatsuda}. Both sets have
been fixed by fitting the masses and decay constants of pseudoscalar
mesons, but it turns out that the HK fit results in somewhat smaller
constituent masses in vacuum. In particular, $M_u^{\mathit vac}
=$~335~MeV which is 33 MeV smaller than the RKH value of 368~MeV used
in Ref.~\cite{BBBNOS02}.  This is still 21~MeV larger than the vacuum
mass in Ref.~\cite{SHH03}.  For the value of the coupling constant in
the diquark channel, $H$, which cannot be determined from meson
properties, we use the same prescription as in Ref.~\cite{BBBNOS02}
for both sets of parameters, i.e., we take the same value as for the
coupling constant in the scalar quark-antiquark channel, $G$. For
comparison, we also consider normal, i.e., color non-superconducting
quark matter, $H=0$. However, we do not consider the possibility of
the intermediate strengths (e.g., $H\simeq 3G/4$) that may give rise
to the recently proposed gapless phases \cite{g2SC,gCFL}.

To describe the hadronic phase we will restrict ourselves to the (relatively
stiff) equation of state derived in Ref.~\cite{xsu3}, which has been
employed in Ref.~\cite{SHH03}. 
The effect of choosing other hadronic equations of state will briefly be
discussed in Sec.~\ref{disc}.

\section{Hybrid equation of state}

By specifying the models for the description of hadron and quark matter, 
we can now determine when the phase transitions between different phases 
happen. We restrict ourselves to the study of 
sharp phase transitions between neutral homogeneous phases of hadronic 
and quark matter. This is partially motivated by the results of
Ref.~\cite{ARRW01} where it was found that a quark-hadron mixed phase
is unlikely to be stable for reasonable values of the surface tension,
and by the findings in Ref.~\cite{NBO03} concerning mixed phases of color
superconducting quark matter.

\begin{figure}
\begin{center}
\epsfig{file=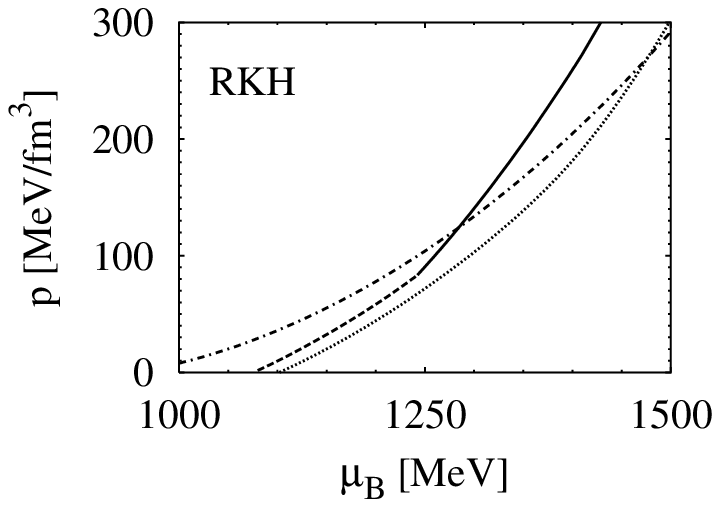,width=7.4cm}
\epsfig{file=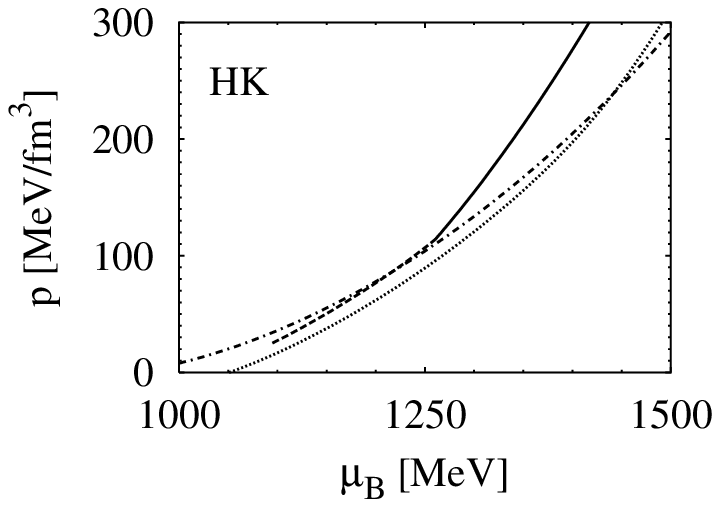,width=7.4cm}
\end{center}
\vspace{-0.5cm}
\caption{\small Pressure of electrically and color neutral matter in beta
equilibrium as a function of baryon chemical potential.
Dash-dotted line: hadronic matter \cite{xsu3}.
The other lines correspond to NJL quark matter in the normal (dotted),
2SC (dashed) or CFL phase (solid), obtained with the parameter sets
RKH~\cite{rehberg} (left panel) or HK~\cite{hatsuda} (right panel).}
\label{figqhphase}
\end{figure}

In Fig.~\ref{figqhphase} the pressure of electrically and color neutral 
hadronic and quark matter in beta equilibrium is displayed as a function 
of the baryon chemical.  
The dash-dotted line corresponds to the hadronic equation 
of state \cite{xsu3}.  
The other lines correspond to NJL quark matter
in the normal phase (dotted), or in a color superconducting
phase. Here we have indicated the part which belongs to the 2SC phase
by a dashed line and the part which belongs to the CFL phase by a
solid line.  Since the two solutions cross each other at the 2SC-CFL
transition point, the slope of the pressure increases discontinuously,
which is clearly visible in the figure.  Physically this is related to
a sudden increase of the baryon number density due to the fact that
the number of strange quarks jumps from almost zero in the 2SC phase
to 1/3 of the total quark number in the CFL phase.  In contrast, there
is no such behavior in the normal quark matter phase where the strange
quarks come in smoothly.

Now, the points of phase transitions are easily read off as the
points of equal pressure, i.e., the points where the lines $p(\mu_B)$
cross.  The results are summarized in Tab.~\ref{tabletransqh}. We
begin our discussion with the left panel of Fig.~\ref{figqhphase}
where the quark equations of state are based on the RKH parameter set.
This is the parameter set used in Ref.~\cite{BBBNOS02}
for the description of quark matter. 

We find that the hadron-quark phase 
transitions occur for both, normal and color superconducting quark 
matter, although, of course, the presence of diquark condensates lowers 
the critical chemical potential substantially. 
In both cases, the phase transition is triggered, to a large extent, 
by the appearance of strange 
quarks in the system. In absence of diquark coupling, the transition 
to normal quark phase happens only at relatively large chemical 
potential that lies well above the strange quark threshold, $\mu_B^{th} 
\simeq 1300$~MeV. At the corresponding critical potential, there are
about 25\% strange quarks. 

In the case of color superconducting quark
matter, a phase transition happens much earlier, leading directly to 
the CFL phase and avoiding an intermediate 2SC phase. This is related 
to the kink in $p(\mu_B)$ at the 2SC-CFL transition point which strongly 
accelerates the phase transition from the hadronic phase. As pointed out 
above, 
the kink in the pressure dependence is associated with the sudden appearance 
of a large amount of strange quarks in the CFL phase. In contrast, 
the diquark condensation in the 2SC color superconducting phase 
practically does not change the slope of the pressure curve. Instead, 
it leads only to a moderate shift of the curve as a whole which can 
be attributed to an additional binding due to the formation of
Cooper pairs. Thus, the use of the RKH parameter set seems to leave
no chance for the presence of the 2SC phase in neutral strongly 
interacting matter. This is in agreement with the conclusions of
Ref.~\cite{BBBNOS02}.


\begin{table}[t]
\begin{center}
\begin{tabular}{|l| c c c c c|}
\hline
&&&&&\\[-3mm]
Phase transition
&  $\mu_B$[MeV] & $\rho_B^{(1)}/\rho_0$ &  $\rho_B^{(2)}/\rho_0$
&  $\epsilon^{(1)}$[MeV/fm$^3$]&  $\epsilon^{(2)}$[MeV/fm$^3$]
\\
&&&&&\\[-3mm]
\hline
&&&&&\\[-3mm]
hadronic\; $\rightarrow$ RKH(CFL) & 1284 & 3.6 & 6.0 &  657 & 1188 \\
hadronic\; $\rightarrow$ RKH(N)   & 1477 & 5.4 & 7.9 & 1092 & 1701\\
&&&&&\\[-2.5mm]
hadronic\; $\rightarrow$ HK(2SC) & 1216 & 3.0 & 3.4 &  536 &  627 \\
HK(2SC)  $\rightarrow$ HK(CFL)   & 1260 & 4.0 & 5.7 &  746 & 1117 \\
hadronic\; $\rightarrow$ HK(N)   & 1441 & 5.1 & 6.5 & 1000 & 1353\\
\hline
\end{tabular}
\end{center}
\caption{\small Various quantities related to the phase transitions
identified in Fig.~\ref{figqhphase}:
critical baryon chemical potential, and baryon and energy densities
below (1) and above (2) the phase transition.}
\label{tabletransqh}
\end{table}  

It turns out, however, that the validity of this statement depends 
strongly on the value of the constituent mass of the non-strange 
quarks in vacuum. As argued above, the latter controls the point 
of zero pressure in the NJL model. Thus, the main
effect of a smaller light quark constituent mass in vacuum is to shift
the pressure curves to lower chemical potentials. This can be seen by
comparing the results on the left hand side of Fig.~\ref{figqhphase}
with the curves displayed in the right panel where we employed the
parameter set HK. There, as a result of the shift, 
we indeed get a phase transition into the 2SC phase
before that is replaced by the CFL phase at a somewhat higher chemical
potential (see Tab.~\ref{tabletransqh} for details). Although the
window is rather small, it makes a qualitative difference for
compact stars, as we will discuss below.

\begin{figure}
\begin{center}
\epsfig{file=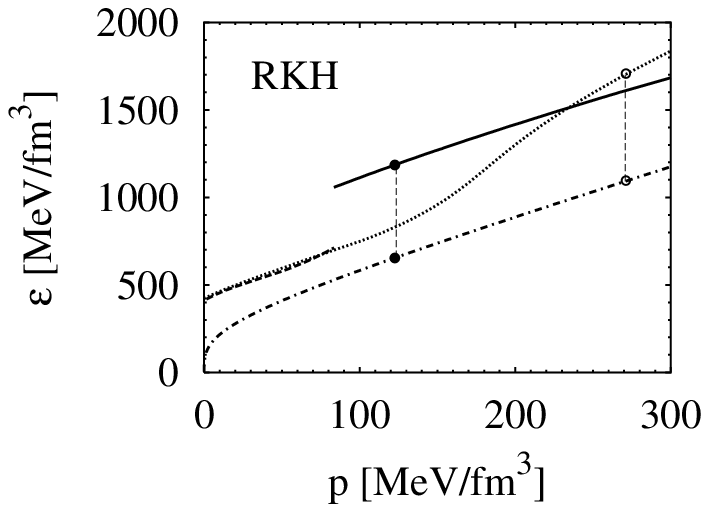,width=7.4cm}
\epsfig{file=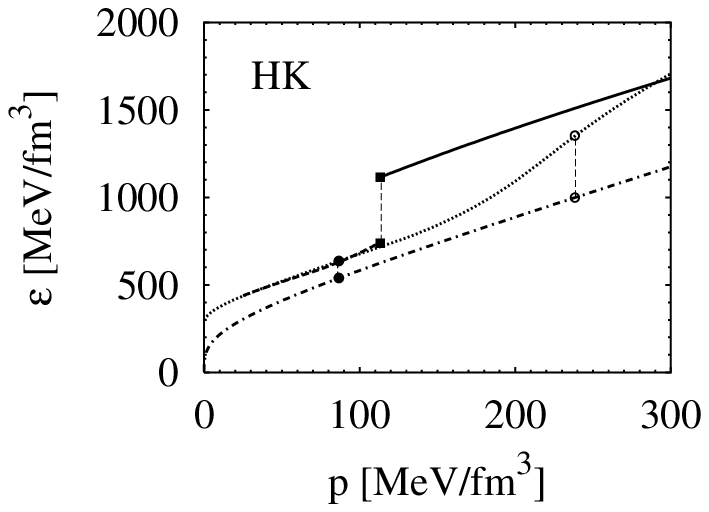,width=7.4cm}
\end{center}
\vspace{-0.5cm}
\caption{\small Energy density as a function of pressure for the hadronic
equation of state~\cite{xsu3} (dash-dotted lines) and 
NJL quark matter obtained with parameter sets RKH~\cite{rehberg} 
(left panel) and HK~\cite{hatsuda} (right panel):
normal quark phase (dotted), 2SC (dashed), CFL (solid). The points with the
thin vertical lines indicate the positions of the phase transitions.}
\label{figeoshq}
\end{figure}
The equations of state, i.e., the energy density as a function of
pressure, for the various phases are presented in Fig.~\ref{figeoshq}.
At the first-order phase transition boundaries the density and, hence,
the energy density increases discontinuously. The corresponding values
are listed in Tab.~\ref{tabletransqh}. As one can see, in most cases
the discontinuity is rather large. The only exception is the
hadron-2SC transition where the jump of the energy density is
relatively small.

\section{Star structure}
\label{starsstars}

For a given equation of state, $\epsilon = \epsilon(p)$, the gravitational
mass $M_G$ of a
static compact star as a function of its radius $R$ can be obtained by
solving the Tolman-Oppenheimer-Volkoff equation~\cite{TOV}.  
\begin{figure}
\begin{center}
\epsfig{file=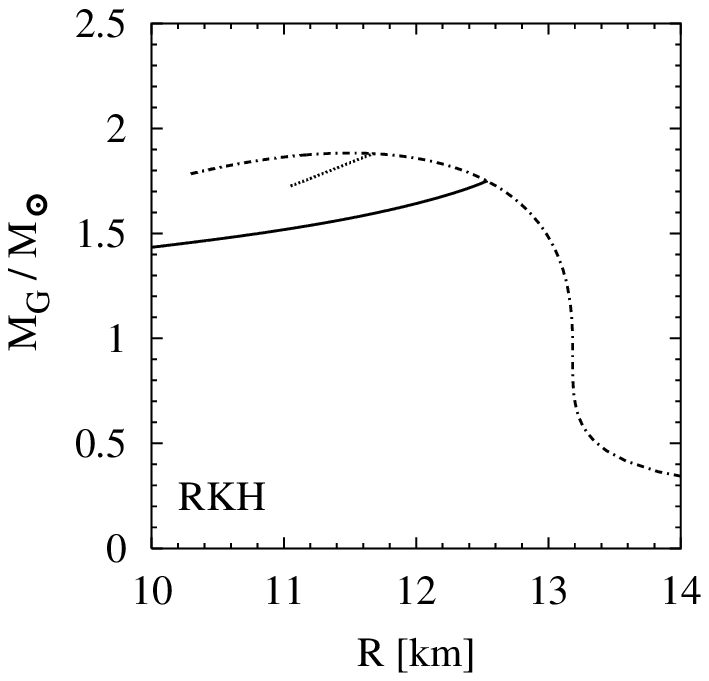,width=7.4cm}
\epsfig{file=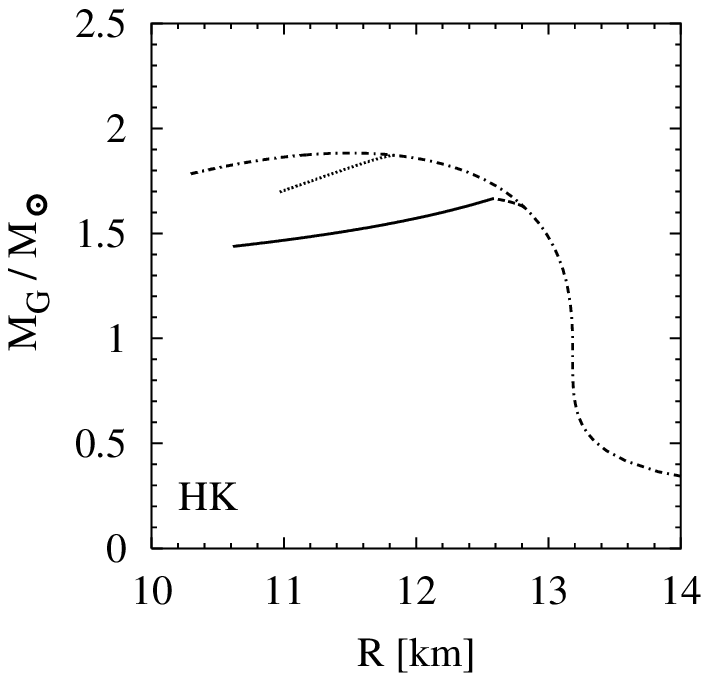,width=7.4cm}
\end{center}
\vspace{-0.5cm}
\caption{\small Gravitational masses of static compact stars as functions 
of the radius for the different equations of state. The dash-dotted lines
indicate the results for a purely hadronic star. 
The other lines indicate
the presence of a quark phase in the center: normal phase (dotted),
2SC dashed, CFL (solid). The quark phases in the left panel have
been calculated with parameter set RKH,
those in the right panel with parameter set HK.}
\label{figtov}
\end{figure}
The resulting curves for the hadronic and
hybrid equations of state constructed above
are displayed in Fig.~\ref{figtov}.
The dash-dotted lines indicate the results for a purely hadronic star.
For a more realistic description of the crust, i.e., the region of
subnuclear matter densities, we have employed the equations of state
of Baym, Pethick and Sutherland~\cite{BPS71} for
$\rho_B<0.001\,\mathrm{fm}^{-3}$ and of Negele and Vautherin~\cite{NV73} for
$0.001\,\mathrm{fm}^{-3} <\rho_B <0.08\,\mathrm{fm}^{-3}$.  The configurations
containing a normal quark matter core are indicated by dotted lines
and those with a color superconducting core in the CFL phase by solid
lines.  As we noted earlier, in the case of the
HK quark equation of state
(right panel), there is a phase transition from hadronic matter to the
2SC phase, followed by a second phase transition to the CFL phase.
Here we have indicated the part of the curve which corresponds to a
2SC phase (but not yet a CFL phase) in the center of the star by a
dashed line. This part of the figure is also presented in an enlarged
form in Fig.~\ref{figtov2}.

\begin{figure}
\begin{center}
\epsfig{file=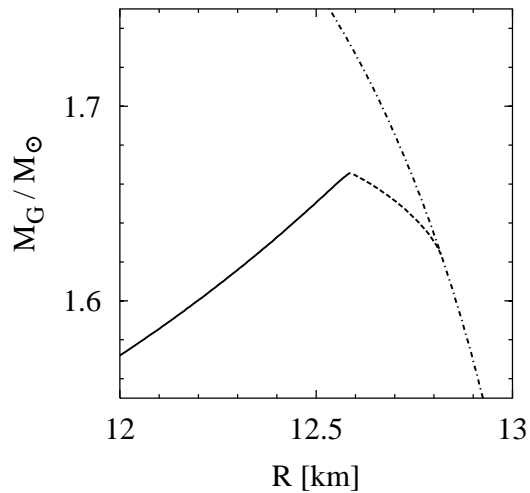,width=7.4cm}
\end{center}
\vspace{-0.5cm}
\caption{\small Enlarged detail of the right panel of Fig.~\ref{figtov},
showing the emergence of a color superconducting quark core in a compact
star calculated with the hybrid equation of state obtained with the 
NJL parameter set HK. 
The meaning of the various line types is the same as in Fig.~\ref{figtov}.}
\label{figtov2}
\end{figure}

The onset of a quark matter phase in the center of the star implies of
course a deviation from the corresponding hadronic matter curve.
Because of the discontinuous energy density at the phase transition
point (see Tab.~\ref{tabletransqh} and Fig.~\ref{figeoshq}) this is
always related to a cusp in the mass-radius relation.  These cusps are
clearly visible in Figs.~\ref{figtov} and \ref{figtov2}.  However, for
all transitions to normal or to CFL quark matter
the effect is so strong that the star is
rendered unstable. This is exactly what has been observed in 
Ref.~\cite{BBBNOS02}.

On the other hand, at the hadron-2SC phase transition
(Fig.~\ref{figtov2}) the star remains stable.
This proofs that stable hybrid stars are possible within the NJL model,
even if strange quarks are included. However, as soon as the 2SC-CFL
phase transition takes place, the star becomes unstable. 
Therefore, in this example, hybrid stars only exist in a small mass window 
between 1.62 and 1.66~$M_\odot$.
Nevertheless, the maximum radius of the 2SC core in a stable configuration
is 3.5 km, about one quarter of the total radius of the star.

Although our assumption of sharp phase transitions between homogeneous
neutral phases seems well justified, since for the transition from
hadronic to quark matter as well as for the transition from the 2SC to
the CFL phase surface energy effects are likely to favor a homogeneous
interface~\cite{ARRW01,NBO03}, we cannot completely exclude the existence
of mixed hadronic-quark phases.
In that case the energy density would not jump, but continuously
interpolate between the hadronic and the quark solution.  As a
consequence, the cusps in Figs.~\ref{figtov} and \ref{figtov2} would
be smoothened out and the instability would not occur immediately at
the onset of the mixed phase.  We expect, however, that in this case
the star would become unstable before the mixed phase turns into a
pure quark phase if this is the CFL or normal quark matter phase.

\section{Discussion}
\label{disc}

In this letter, we continued the systematic study of 
Refs.~\cite{AlRe03,BB02,BBBNOS02,BGAYT,SHH03} in search of possible 
constructions of hybrid stars with color superconducting quark 
matter in their interior. In order to understand the striking 
differences in the conclusions of Ref.~\cite{BBBNOS02}, where no
stable hybrid star constructions were found, and Ref.~\cite{SHH03},
where the hybrid stars with the 2SC quark cores were the most natural
constructions, we considered a combination of a specific 
hadronic equation of state with two different equations of state 
for quark matter. Without much of a limitation, we used only 
constructions with sharp boundaries between different phases of 
matter.

While the hadronic equation of state is without any doubt a very 
important input in the constructions of hybrid baryonic matter,
here we concentrate mostly on the role of the quark equations 
of state. 
In bag models, like in Ref.~\cite{AlRe03}, these are mainly
determined by the values of the bag constant, the diquark gap, and 
the strange quark mass. 
In NJL-type models, the quarks acquire effective (``constituent'') masses 
through
dynamical breaking of chiral symmetry. Although this is in some way related
to an effective bag constant, the effect is more complex and cannot be
described by a single density independent number.
In particular, the strange and the non-strange sectors partially decouple.

In the present letter we show that a very important
parameter in NJL-type models of quark matter is the value of
the constituent mass of the {\it non-strange} quarks in vacuum. This
parameter controls the value of the baryon chemical potential at 
which the pressure is zero. For example, decreasing 
(increasing) the mass parameter has the effect of shifting the 
whole dependence of the pressure to smaller (larger) values of 
the baryon chemical potential and thereby the 
hadron-quark phase transition to lower (higher) densities.
Therefore, for relatively small values of the mass,
the first quark phase that appears with increasing baryon 
density is the two-flavor color superconducting (2SC) phase 
which is eventually replaced by the color-flavor locked (CFL) phase 
at higher densities. For large 
values of the mass, on the other hand, the phase transition 
goes from the hadronic phase directly into the CFL phase 
avoiding the 2SC phase. 

This difference turned out to be crucial for the stability of hybrid stars.
We found that hybrid stars, if they exist, have 
a quark core in the 2SC phase and contain only a small fraction of 
strange quarks. 
Also, our study suggests that the size of a quark matter 
core inside a hybrid star is very sensitive to the choice
of the non-strange quark mass parameter in the NJL model.
In particular, when one goes from the ``small" mass limit,
as in Ref.~\cite{SHH03}, to the ``large" mass limit,
as in Ref.~\cite{BBBNOS02}, the quark core radius changes 
from being very large to vanishing. 

In the NJL model studied here, we find no stable stars with either 
CFL or normal quark matter cores. This is in contrast to the prediction of 
Ref.~\cite{AR02} where it was argued that there is no 2SC phase in 
compact stars. 
Let us be more precise: 
Performing a Taylor expansion in the strange quark mass, the authors 
of Ref.~\cite{AR02} found that in beta-equilibrated electrically and
color neutral quark matter the 2SC phase is always less favored than 
the CFL phase or normal quark matter. From this observation they 
concluded that the 2SC phase is absent in compact stars.
In contrast to this result, it was shown in Ref.~\cite{SRP02} in the
framework of the NJL model that neutral 2SC matter could be the most 
favored quark phase in a certain regime. 
However, the authors argued that this interval
might disappear if the hadronic phase is included more properly.
This is indeed what we found for parameter set RKH, while for parameter
set HK the 2SC phase survives only in a tiny window. 
Nevertheless, if Nature chooses to be similar to this equation of state,
it will be this tiny window which gives rise to hybrid stars, whereas 
the CFL phase would be never present in compact stars. 

At this point we should ask, to what extent our results, which are
based on two examples can be considered as a general feature.  
To address this, let us briefly come back to the dependence on the hadronic
equation of state.
As mentioned earlier, in Ref.~\cite{BBBNOS02} three different hadronic
equations of state have been considered.
For the stiffest one, obtained within a microscopic
Brueckner-Hartree-Fock calculation without hyperons~\cite{bbb},
the results are in qualitative agreement with ours
if the quark phase is described by the RKH equation of state.
The two other hadronic  equations of state employed in Ref.~\cite{BBBNOS02}
are softer, thus rendering a transition to quark matter less favorable.
In fact, for the softest equation of state, based on a
Brueckner-Hartree-Fock calculation with hyperons~\cite{bbs}, 
no phase transition was found at all, while for the intermediate one,
based on a relativistic mean-field model~\cite{Glendenning}, a
phase transition took place only if diquark pairing was taken into
account.
Employing these three hadronic equations of state, we found that none
of the results reported in Ref.~\cite{BBBNOS02} changes qualitatively
if the NJL parameters HK are used instead of RKH.
In particular, in none of these cases there is an intermediate 2SC phase
and there are no stable hybrid stars.

Both, an intermediate 2SC phase and stable hybrid stars,
thus appear as rather exceptional cases which we only obtained
if the very stiff hadronic equation of state based on Ref.~\cite{xsu3}
is combined with the HK quark equation of state. The constituent mass
of the non-strange quarks thereby plays a key role.
Note in addition that our choice $H=G$ for the diquark coupling constant is
likely to be an upper limit.  With a smaller coupling,
there would be less binding in the 2SC phase and eventually the small
window of stable hybrid stars which is seen in Fig.~\ref{figtov2}
may close.
On the other hand, if we further decrease the constituent mass of the
non-strange quarks, the window could become wider. Similarly,
a larger constituent mass of the strange quarks would render the CFL
phase less favored and thereby help stabilizing the 2SC phase. 
A systematic study in all these directions still remains to 
be done. 
This should also include the consideration of other color superconducting
phases which have not been taken into account here.

Whereas the question about details of the hadron-quark phase transition
is thus strongly dependent on both, the hadronic equation of state
and the choice of the parameters for the quark phase,
all NJL-model calculations so far agree that there is no stable hybrid
star with a quark core in the CFL phase or in the normal phase with
a large fraction of strange quarks. 
As mentioned earlier, this observation, which is in contrast to the 
bag-model results of Ref.~\cite{AlRe03}, can be explained by the fact
that the dynamically generated 
effective bag constants and effective strange quark masses
of the NJL model are in general quite large.
Also, the fact that these effective quantities depend on the phase
and are therefore different for normal, 2SC and CFL quark matter
leads to important differences to the bag model (see Ref.~\cite{BBBNOS02}).

Finally, it should be reminded that our results rely on the
assumption that the NJL model parameters which have been fitted in
vacuum can be applied to dense matter. It is of course possible that
this is not the case.  Hence our arguments could also be turned around. 
The observation, e.g., of a compact star with a CFL quark matter
core or even a pure strange quark star would convincingly demonstrate
that the quark-matter phase is not well described by an NJL-type model
with parameters which are fixed in vacuum.

\section*{Acknowledgments}
We would like to thank M. Hanauske for providing us
with the data of their hadronic equation of state~\cite{xsu3}.  
We also thank V. Werth for having pointed out to us a numerical mistake
in the first version of the manuscript.
The work of I. S. was supported by
Gesellschaft f\"{u}r Schwerionenforschung (GSI) and by
Bundesministerium f\"{u}r Bildung und Forschung (BMBF).



\begin{thebibliography}{99}
\bibitem{ARW98}
M.G.~Alford, K.~Rajagopal and F.~Wilczek,
Phys.\ Lett.\ B {\bf 422} (1998) 247.

\bibitem{RSSV98} 
R.~Rapp, T.~Sch\"afer, E.V.~Shuryak and M.~Velkovsky,
Phys.\ Rev.\ Lett.\  {\bf 81} (1998) 53.

\bibitem{RaWi00} 
K.~Rajagopal and F.~Wilczek,
hep-ph/0011333 and references therein.

\bibitem{Alford01}
M.G.~Alford,
Ann.\ Rev.\ Nucl.\ Part.\ Sci.\  {\bf 51} (2001) 131.

\bibitem{Schaefer03} 
T.~Sch\"afer, 
hep-ph/0304281.

\bibitem{Rischke03} 
D.H.~Rischke, 
nucl-th/0305030.

\bibitem{AlRe03} 
M.~Alford and S.~Reddy,
Phys.\ Rev.\ D {\bf 67} (2003) 074024.


\bibitem{BB02} S.~Banik and D.~Bandyopadhyay,
Phys.\ Rev.\ D {\bf 67} (2003) 123003.


\bibitem{BBBNOS02} 
M.~Baldo, M.~Buballa, F.~Burgio, F.~Neumann, M.~Oertel and H.J.~Schulze,
Phys.\ Lett.\ B {\bf 562} (2003) 153.

\bibitem{BGAYT} 
D.~Blaschke, H.~Grigorian, D.N.~Aguilera, S.~Yasui and H.~Toki,
AIP Conf.\ Proc.\  {\bf 660} (2003) 209.

\bibitem{SHH03}
I.~Shovkovy, M.~Hanauske and M.~Huang,
Phys.\ Rev.\ D {\bf 67} (2003) 103004.

\bibitem{LH02} 
G.~Lugones and J.~E.~Horvath,
Phys.\ Rev.\ D {\bf 66}, 074017 (2002)
[arXiv:hep-ph/0211070].

\bibitem{SLS99}
K.~Schertler, S.~Leupold and J.~Schaffner-Bielich,
Phys.\ Rev.\ C {\bf 60} (1999) 025801.

\bibitem{SPL00} 
A.~Steiner, M.~Prakash and J.M.~Lattimer,
Phys.\ Lett.\ B {\bf 486} (2000) 239.

\bibitem{AR02}M.~Alford and K.~Rajagopal,
JHEP {\bf 0206} (2002) 031.

\bibitem{SRP02}
A.~W.~Steiner, S.~Reddy and M.~Prakash,
Phys.\ Rev.\ D {\bf 66}, 094007 (2002)
[arXiv:hep-ph/0205201].

\bibitem{NBO03}
F.~Neumann, M.~Buballa and M.~Oertel,
Nucl.\ Phys.\ A {\bf 714} (2003) 481.

\bibitem{rehberg}
P.~Rehberg, S.P.~Klevansky and J.~H\"ufner,
Phys.\ Rev.\ C {\bf 53} (1996) 410.

\bibitem{hatsuda}
T.~Hatsuda and T.~Kunihiro,
Phys.\ Rept.\  {\bf 247} (1994) 221.

\bibitem{g2SC}
I.~Shovkovy and M.~Huang,
Phys.\ Lett.\ B {\bf 564} (2003) 205;
Nucl.\ Phys.\ A {\bf 729} (2003) 835.

\bibitem{gCFL}
M.~Alford, C.~Kouvaris and K.~Rajagopal,
hep-ph/0311286.

\bibitem{xsu3} M.~Hanauske, D.~Zschiesche, S.~Pal, S.~Schramm,
H.~St\"ocker and W.~Greiner,
Astrophys.\ J.\  {\bf 537}, 50320 (2000).

\bibitem{ARRW01}
M.G.~Alford, K.~Rajagopal, S.~Reddy and F.~Wilczek,
Phys.\ Rev.\ D {\bf 64} (2001) 074017.

\bibitem{TOV} 
R.C.~Tolman,
Phys.\ Rev.\  {\bf 55}, 364 (1939);
J.R.~Oppenheimer and G.M.~Volkoff,
Phys.\ Rev.\  {\bf 55}, 374 (1939).

\bibitem{BPS71} G.~Baym, C.~Pethick and P.~Sutherland,
Astrophys.\ J.\  {\bf 170}, 299 (1971).

\bibitem{NV73} J. Negele and D. Vautherin,
Nucl.\ Phys.\ {\bf A207}, 298 (1973).

\bibitem{bbb}
M.~Baldo, I.~Bombaci and G.F.~Burgio, 
Astron.\ Astrophys.\ {\bf 328} (1997) 274.

\bibitem{bbs}
M.~Baldo, G.F.~Burgio and H.-J.~Schulze,
Phys.\ Rev.\ C {\bf 58} (1998) 3688;\\
Phys.\ Rev.\ C {\bf 61} (2000) 055801.

\bibitem{Glendenning} N.K. Glendenning, {\sl Compact Stars} 
(Springer, New York, 1996).


\end{thebibliography}
\end{document}